\documentclass{egpubl}
\usepackage{eurovis2023}

\SpecialIssuePaper         

\CGFStandardLicense

\usepackage[T1]{fontenc}
\usepackage{dfadobe}  

\usepackage{cite}  
\BibtexOrBiblatex
\electronicVersion
\PrintedOrElectronic
\ifpdf \usepackage[pdftex]{graphicx} \pdfcompresslevel=9
\else \usepackage[dvips]{graphicx} \fi

\usepackage{egweblnk}

 \usepackage{multirow}
 \usepackage[table,xcdraw]{xcolor}
 \usepackage{amsmath}
 \usepackage{hyperref}
 \usepackage{enumitem}
 \usepackage[export]{adjustbox}
 \usepackage{arydshln}
 \usepackage{dblfloatfix}
 \usepackage{booktabs}
 \usepackage{array}
 \usepackage{tabularx}
 \usepackage{soul}
 \usepackage{url}

\usepackage{xcolor}

\title[Mini-VLAT: A Short and Effective Measure of Visualization Literacy]%
      {Mini-VLAT: A Short and Effective Measure of \\Visualization Literacy}
      
\author[S. Pandey \& A. Ottley]
{\parbox{\textwidth}{\centering Saugat Pandey\orcid{0000-0002-7429-6575}
        and Alvitta Ottley\orcid{0000-0002-9485-276X} 
        }
        \\
{\parbox{\textwidth}{\centering Washington University in St.\ Louis, St.\ Louis, MO, USA
       }
}
}

\begin{document}

\maketitle
\begin{abstract}
The visualization community regards visualization literacy as a necessary skill. Yet, despite the recent increase in research into visualization literacy by the education and visualization communities, we lack practical and time-effective instruments for the widespread measurements of people's comprehension and interpretation of visual designs. We present Mini-VLAT, a brief but practical visualization literacy test. The Mini-VLAT is a 12-item short form of the 53-item Visualization Literacy Assessment Test (VLAT). The Mini-VLAT is reliable (coefficient omega = 0.72) and strongly correlates with the VLAT. Five visualization experts validated the Mini-VLAT items, yielding an average content validity ratio (CVR) of 0.6. We further validate  Mini-VLAT by demonstrating a strong positive correlation between study participants' Mini-VLAT scores and their aptitude for learning an unfamiliar visualization using a Parallel Coordinate Plot test. Overall, the Mini-VLAT items showed a similar pattern of validity and reliability as the 53-item VLAT. The results show that Mini-VLAT is a psychometrically sound and practical short measure of visualization literacy.


\end{abstract}  
\section{INTRODUCTION}

The proliferation of visualization technologies, toolkits, journalism, and social media has led to the pervasiveness of visualization in people’s daily lives. However, people’s capacity to interpret and use graphics varies~\cite{peck2012towards,ottley2020adaptive,liu2020survey}, leaving the visualization community with unsolved questions with far-reaching consequences. With this increased use of visualization tools to convey complex data, measuring visualization literacy (VL) is becoming increasingly vital. 

VL is a crucial skill for understanding and interpreting complex imagery communicated by interactive visual designs \cite{bach2021special}.
As a result, researchers in the Information Visualization community have attempted to examine users’ visualization literacy \cite{joshivisualization, borner2019data, boy2014principled, peppler2021cultivating,carswell1992choosing}. Across several studies, scholars have proposed methods to assess visualization literacy, attempted to understand the current state of data visualization comprehension among the general public~\cite{peck2019data,borner2019data}, and suggested various ways of learning unfamiliar visualizations to improve users’ visualization literacy. 

Although VL has been described as an essential skill through various workshops, keynote talks, and research works \cite{bach2021special, joshivisualization}, researchers still lack a validated and accurate tool for testing user visualization literacy that can be widely deployed \cite{yi2012implications}. Perhaps the most well-known publicly available visualization literacy test, Visualization Literacy Assessment Test (VLAT) \cite{lee2016vlat}, consists of 53 questions across 12 different visualization types. Each question has a limit of 25 seconds, making it approximately 22 minutes long. Such a long test typically necessitates strong cognitive demands, which might lead to cognitive tiredness and erroneous answers for some respondents \cite{sitarenios2022short}. Thus, the time commitment involved with VLAT limits its use to primarily academic settings. 

The recent discussions about visualization literacy at IEEE VIS 2022 focused on developing methods to test visualization literacy and an appropriate alternative for VLAT. One of the issues raised was the need for a short and effective visualization literacy test that can be used for large-scale data collection and visualization evaluation. 
 \textit{To address this need, we developed an abbreviated version of the Visualization Literacy Assessment Test, Mini-VLAT.} The Mini-VLAT consists of 12 multiple-choice questions spread across 12 different visualization types, making it 5 minutes long. We evaluated the Mini-VLAT with the established procedure of test development in Psychology and Established Measurement \cite{cohen1996psychological} and other highly cited works in the field of psychology \cite{sitarenios2022short, smith2000sins} when creating a shortened test in 5 phases: (1) Test Item Generation (2) Piloting and Item Refinement (3) Reliability Evaluation (4) Correlation between Mini-VLAT and VLAT. (5) Testing Mini-VLAT's predictive capacity.

In summary, this paper presents Mini-VLAT, a brief and practical test for measuring visualization literacy that can be widely used. We follow established procedures for test development and validation, including test generation, reliability evaluation, correlation analysis with the established VLAT, and testing the predictive capacity of Mini-VLAT. Our results demonstrate that Mini-VLAT is a psychometrically sound and reliable measure of visualization literacy that is strongly correlated with VLAT and has similar validity and reliability properties. The online version of Mini-VLAT can be found at \url{https://washuvis.github.io/minivlat/}.

\section{RELATED WORK}

Substantial work has been done around defining and examining visualization literacy. In this section, we discuss previous works on defining visualization literacy and then discuss relevant works that examined visualization literacy. 

\subsection{Definitions of Visualization Literacy}
Several seminars at visualization conferences have sparked conversations among scholars to define visualization literacy. It intersects with several disciplines of study, including cognitive psychology and education. Although several researchers have attempted to define visualization literacy, there is no clear consensus. 

In an early study, graph literacy was used instead of visualization literacy \cite{galesic2011graph}. Galesic et al.\cite{galesic2011graph} created a test scale to assess graph literacy in the context of health. The test items were divided into three comprehension levels: reading the data, reading between the data, and reading beyond the data.
However, Boy et al. \cite{boy2014principled} laid the groundwork for visualization literacy in the Information Visualization community. They defined visualization literacy as 
\begin{quote}
    "The ability to confidently use a given data visualization to translate questions specified in the data domain into visual queries in the visual domain, as well as interpreting visual patterns in the visual domain as properties in the data domain." \cite{boy2014principled}
\end{quote}
 In 2015, Börner et al. \cite{borner2016investigating} investigated the familiarity of different visualization types among youth and adult museum visitors. They defined visualization literacy as 
 \begin{quote}
     "The ability to make meaning from and interpret patterns, trends, and correlations in visual representations of data." \cite{borner2016investigating}
\end{quote}

Based on these definitions, Lee et al. \cite{lee2016vlat} devised their own definition of visualization literacy and introduced \textit{Visualization Literacy Assessment Test (VLAT)}. They defined visualization literacy as
 \begin{quote}
     "the ability and skills to read and interpret visually represented data in and to extract information from data visualizations." \cite{lee2016vlat}
\end{quote}

We assert that visualization literacy is a multidimensional construct, and measuring its full scope with a single scale is untenable. Thus, this work focuses on \textit{measuring one's ability to read and interpret visually represented data}, which we argue is a necessary skill for every proposed definition of VL.
\subsection{Measuring Visualization Literacy}

Several research works have been proposed towards understanding and measuring visualization literacy \cite{galesic2011graph, firat2022p, firat2020treemap, lee2016vlat, koedinger2001toward, borner2016investigating}. Most early works have measured the ability to read and interpret a graph \cite{galesic2011graph, bertin2010semiology, carswell1992choosing, wainer1992understanding}. However, the question items in these works consisted of primitive graphs based on the three comprehension levels mentioned before. 

Differing from this prior work, Boy et al. \cite{boy2014principled} applied item response theory to measure visualization literacy. The items in the test were based on six visualization tasks (minimum, maximum, variation, intersection, average, and comparison). The test included line charts, bar charts, and scatterplots. 

Börner et al. \cite{borner2016investigating} applied yet another approach to measuring data visualization literacy by examining the familiarity of various visualization designs.  
Three US science museums hosted this investigation. Börner et al. \cite{borner2016investigating} selected 20 visualizations from textbooks and popular web visualization libraries, such as the D3.js library. The selected visualizations contain two charts, five maps, eight graphs, and five network layouts. Five of the twenty visual designs were shown to visitors, who were asked to express their knowledge of the visual designs and identify the design's name.

Although this work does not attempt to create a visualization literacy measure, it does provide insight into how people had significant limitations in identifying and understanding different data visualizations. Subsequently, Börner et al. \cite{borner2019data} published a visualization literacy framework paper that included a set of definitions, conceptual frameworks, exercises, and assessments to describe visualization literacy. They introduced a data visualization literacy framework (DVL-FM) based on seven core hierarchical typologies of extracting insights from data \cite{borner2019data}.

In 2016, Lee et al. \cite{lee2016vlat} developed a visualization literacy assessment test (VLAT) that consists of 12 data visualizations and 53 multiple-choice test items. This test demonstrated high reliability and validity. The test was also validated among five visualization experts. The VLAT encouraged researchers to extend the work by adding more advanced visual designs \cite{firat2020treemap, firat2022p}. Firat et al. \cite{firat2022p} developed a parallel coordinates literacy test, P-Lite, with diverse images generated using popular PCP software tools. Like VLAT, P-Lite was created to test the user's comprehension and ability to interpret high-dimensional visual designs like parallel coordinates. Firat et al. \cite{firat2020treemap} also created a treemap literacy test to assess the user's understanding and ability to interpret treemaps. 

However, the length of the VLAT serves as its limitation. It limits the researchers in the visualization community to investigate the users' ability to read and interpret visually represented data on a larger scale. In this study, we present an abbreviated version of the large VLAT that has comparable reliability and validity as the VLAT with 12 items.

\section{WHY 12-ITEM Mini-VLAT?}
Underscoring the need for a shorter visualization literacy test, some works have attempted to shorten the VLAT for research purposes \cite{mansoorlinking, lee2019correlation}. However, none of them followed the guidelines concerning the development of short forms \cite{smith2000sins, stanton2002issues}. For example, Lee et al. \cite{lee2019correlation} reduced the number of items from 53 to 41 to demonstrate the correlation between the users' cognitive characteristics and visualization literacy. They shortened that test by removing the 12 items with the lowest discriminating values, i.e., the items that do not reliably distinguish the high and low performers on the test.
Mansoor et al. \cite{mansoorlinking} also decreased the number of items from 53 to 22, citing the participants' performance degradation over time due to fatigue in their pilot study. However, a 22-item or 41-item test could still be too long for some studies.

What distinguishes VLAT from other visualization literacy assessments is how comprehensive it is by testing 12 different visualization designs with up to 7 types of questions for each. Inevitably, a short form will produce a less complete test, and we weighed the tradeoff between maintaining the variety of visualization designs or the task types. To create an abbreviated test that can be used in a wide range of non-experts settings, we opted to maintain the diversity of the visualization types.
We hypothesize that we could produce a reliable and valid abbreviated test by selecting only one item from each visualization type if done carefully. It will also significantly decrease the completion time, making it suitable for widespread use when measuring visualization literacy. In the following sections, we describe the Mini-VLAT development process in detail. 

\section{DEVELOPMENT OF THE Mini-VLAT}
The literature review demonstrates several existing methods for examining people's ability to read, interpret, and extract information from data visualization. However, we assert that a short but valid measure will be more practical for widespread use. Ultimately, we aim to develop a tool for researchers and visualization designers to gauge whether people can read visualization and track how visualization literacy changes over time. To achieve this, we opted to shorten an existing survey and leverage best practices in psychology to develop short-form surveys \cite{smith2000sins, sitarenios2022short, donnellan2006mini, melzack1987short}. 

\noindent
We accomplish this through a 5-phase process: 
\begin{enumerate}[label=\textbf{Phase
\Roman*:},leftmargin=4\parindent]
    \item We began by replicating the VLAT and computing the total-item correlation for each item in the test. We then invited five domain experts in the Information Visualization community to evaluate the items for the large Mini-VLAT \cite{american1974standards, mislevy2006implications}. We selected the item with high item-total correlation and positive \textit{CVR} from each visualization type for the Mini-VLAT.
    \item The items selected in Phase I were tried on a sample of test takers. Based on the results from the analysis, we made revisions to the questions and included them in the final set of the Mini-VLAT.
     \item The items in the Mini-VLAT were evaluated in terms of reliability. We used the most commonly used internal consistency reliability measure called \textit{reliability coefficient omega ($\omega$)}. The reliability evaluation's findings provide additional validation for the validity that visualization literacy is measured accurately and consistently \cite{american1974standards}.
     \item It is critical to demonstrate sufficient variation between the short and full forms. Research studies generally calculate the correlation between the short and long forms based on a single test administration \cite{canivez1998long, lafreniere1996social}. Smith et al. \cite{smith2000sins} called this a methodological error because it leads to an overestimation of the correlation between the two forms. The answers to every question on the short form are counted twice, appearing on both sides of the correlation. So, naturally, any error or random variation in the responses to any of the short-form items is completely replicated in the long--form. By using this method, one is in fact correlating error to itself to some extent. In this phase, we compared the scores in the large Mini-VLAT and VLAT using Pearson’s product-moment correlation coefficient. We established a strong positive correlation between the two forms.
     \item For our final validation test, we test how well the scores on the Mini-VLAT predict the performance on an independent aptitude test. For the aptitude test, we used the items from the Parallel Coordinate Test, P-Lite \cite{firat2022p}.
\end{enumerate}
In the following sections, we describe the five stages of construction and evaluate the large Mini-VLAT.

\begin{table*}[h]
\scriptsize
\centering
\caption{The test items in the VLAT and the item-total correlation (\textit{r})  and content validity ratio (\textit{CVR}) of each item. An item with {\color[HTML]{3182BD} highest} item-total correlation for each visualization type. One {\color[HTML]{F19976} item} from each visualization type initially selected for the Mini-VLAT. Comparison between the responses from the {\color[HTML]{009966}original} and {\color[HTML]{FF6600}replicated} study for each item using error bars (95\% CI).}
\label{tab:vlat-rep}
\renewcommand{\arraystretch}{1}%

\begin{tabular*}{\textwidth}{|cclcc|p{3.15cm}|}
\hline

\multicolumn{1}{|c}{\textbf{Item ID}} & \textbf{Visualization Type} & \multicolumn{1}{c}{\textbf{Question}} & \multicolumn{1}{c}{\textbf{r}} & \multicolumn{1}{c|}{\textbf{CVR}} &   \textbf{ {\color[HTML]{009966}Original} vs. {\color[HTML]{FF6600}Replicated} VLAT } \\ \hline 

%
%
 &  &  &  &  &  \\ [-2ex]
1 &  &  {\color[HTML]{F19976} What was the price of a barrel of oil in February 2015?} & 0.43 & 0.2 & \\ 
2 &  & In which month was the price of a barrel of oil the lowest in 2015? & 0.15 & 0.2 & \\
3 &  & What was the price range of a barrel of oil in 2015? & {\color[HTML]{3182BD} 0.59} & -0.2 &  \\
4 &  & Over the course of the second half of 2015, the price of a barrel of oil was \_\_\_\_\_\_\_\_\_\_\_\_. & 0.22 & -0.2 &  \\
5 & \multirow{-5}{*}{Line Chart} & \begin{tabular}[c]{@{}l@{}}About how much did the price of a barrel of oil fall from April to September in 2015?\end{tabular} & 0.36 & 0.2 & {\multirow{-5}{*}{\raisebox{10pt}[\dimexpr\height-.5\baselineskip\relax]%
\centering
\hspace*{-.2cm}%
{\includegraphics[width=3.5cm]{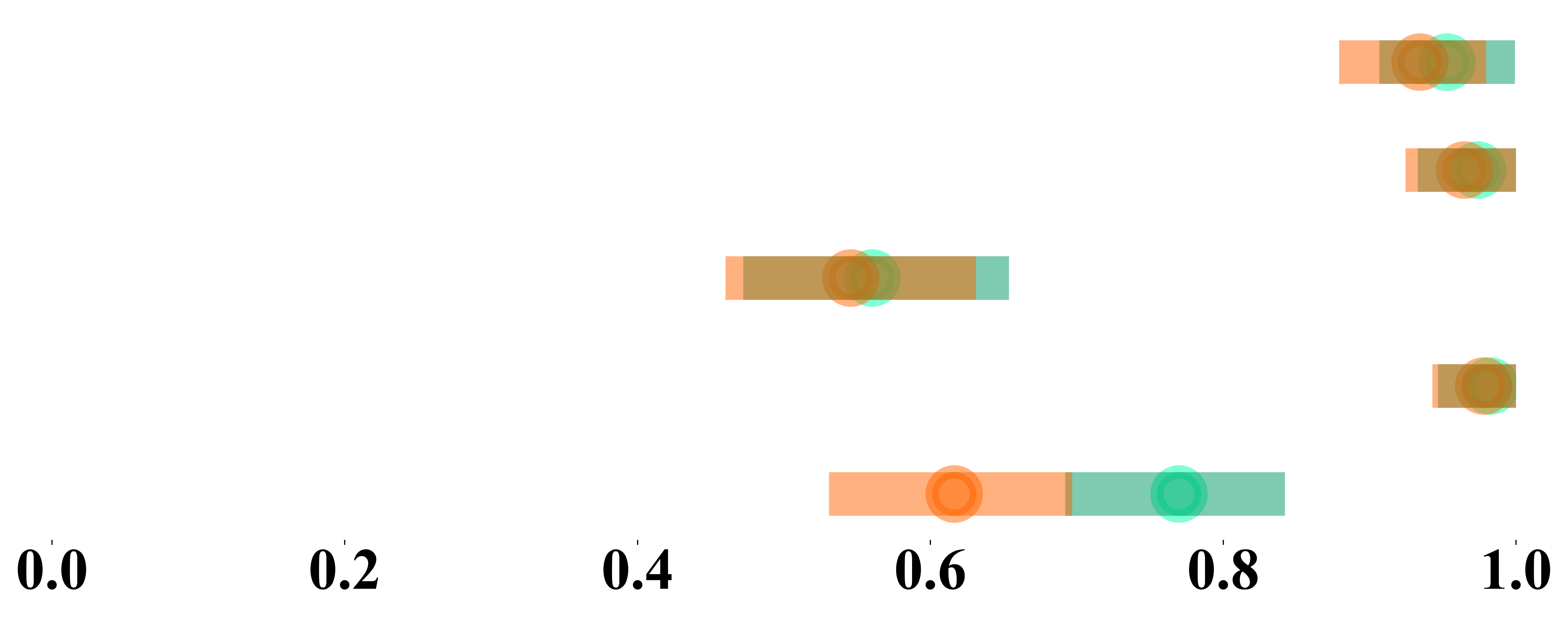}
\hspace*{-3cm}%
}}} \\
 &  &  &  &  &  \\ [-1ex]
\hline

%
%
 &  &  &  &  &  \\ [-2ex]
6 &  & {\color[HTML]{F19976} What is the average internet speed in Japan?} & 0.29 & 0.6 &  \\
7 &  & In which country is the average internet speed the fastest in Asia? & 0.12 & 0.6 &  \\
8 &  & What is the range of the average internet speed in Asia? & {\color[HTML]{3182BD} 0.47} & -0.6 &  \\
9 & \multirow{-4}{*}{Bar Chart} & How many countries in Asia is the average internet speed slower than Thailand? & 0.03 & -0.2 & {\multirow{-4}{*}{\raisebox{0pt}[\dimexpr\height-0.5\baselineskip\relax]%
\centering
\hspace*{-.2cm}%
{\includegraphics[width=3.5cm]{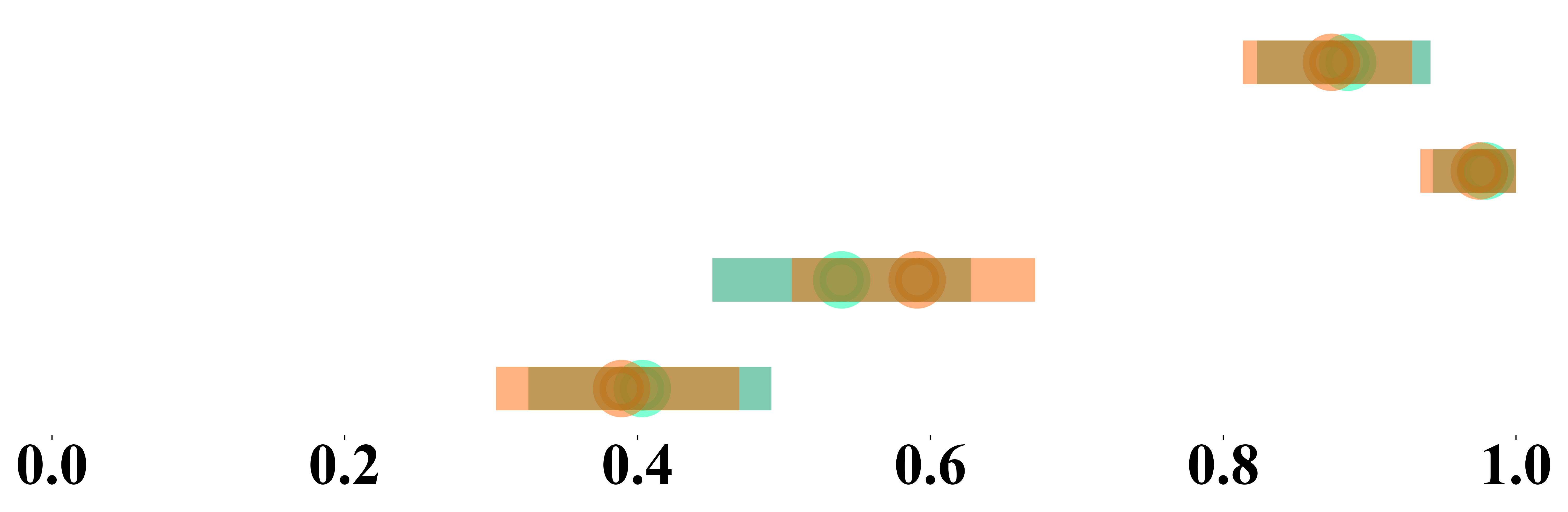}
\hspace*{-3cm}%
}}} \\
 &  &  &  &  &  \\ [-1ex]
 \hline

%
%
 &  &  &  &  &  \\ [-2ex]
10 &  & {\color[HTML]{F19976} What is the cost of peanuts in Las Vegas?} & 0.33 & 0.2 &  \\
11 &  & About what is the ratio of the cost of a sandwich to the total cost of room service in Seattle? & 0.26 & 0.6 &  \\
12 &  & In which city is the cost of soda the highest? & {\color[HTML]{3182BD} 0.37} & -0.6 &  \\
13 &  & The cost of vodka in Atlanta is higher than that of Honolulu. & 0.27 & 0.6 &  \\
14 & \multirow{-5}{*}{Stacked Bar Chart} & \begin{tabular}[c]{@{}l@{}}The ratio of the cost of Soda to the cost of Water in Orlando is higher than that of Washington D.C.\end{tabular} & 0.23 & -0.2 & {\multirow{-5}{*}{\raisebox{0pt}[\dimexpr\height-0.5\baselineskip\relax]%
\centering
\hspace*{-.2cm}%
{\includegraphics[width=3.5cm]{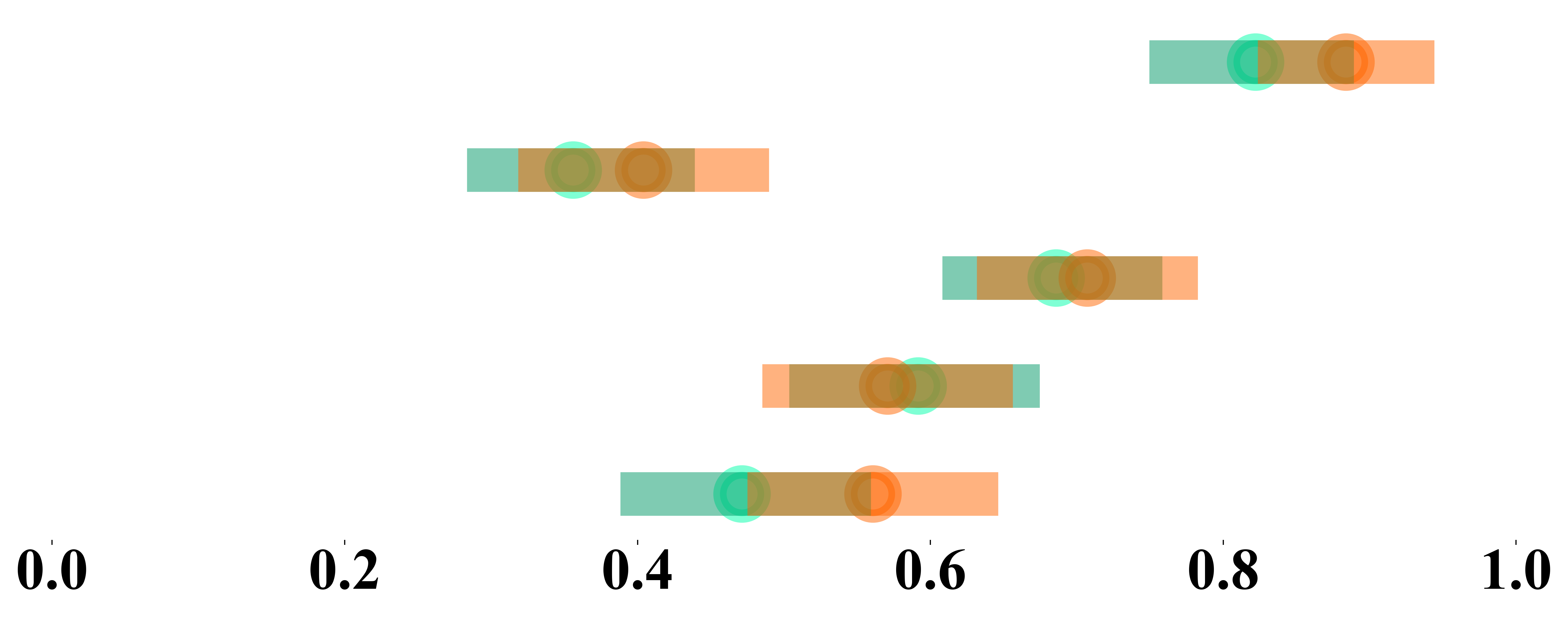}
\hspace*{-3cm}%
}}} \\ 
 &  &  &  &  &  \\ [-1ex]
\hline

%
%
&  &  &  &  &  \\ [-2ex]
15 &  & \begin{tabular}[c]{@{}l@{}}What is the approval rating of Republicans among the people who have the education \\ level of Postgraduate Study?\end{tabular} & 0.35 & 0.2 &  \\
16 &  & {\color[HTML]{F19976} What is the education level of people in which the Democrats have the lowest approval rating?} & {\color[HTML]{3182BD} 0.37} & 1 &  \\
17 & \multirow{-3}{*}{100\% Stacked Bar Chart} & \begin{tabular}[c]{@{}l@{}}The approval rating of Republicans for the people who have the education level of Some  College \\Degree is lower than that for the people who have the education  level of Postgraduate Study.\end{tabular} & 0.20 & 0.6 & \multirow{-4}{*}{\raisebox{0pt}[\dimexpr\height-0\baselineskip\relax]%
\centering
\hspace*{-.2cm}%
{\includegraphics[width=3.5cm]{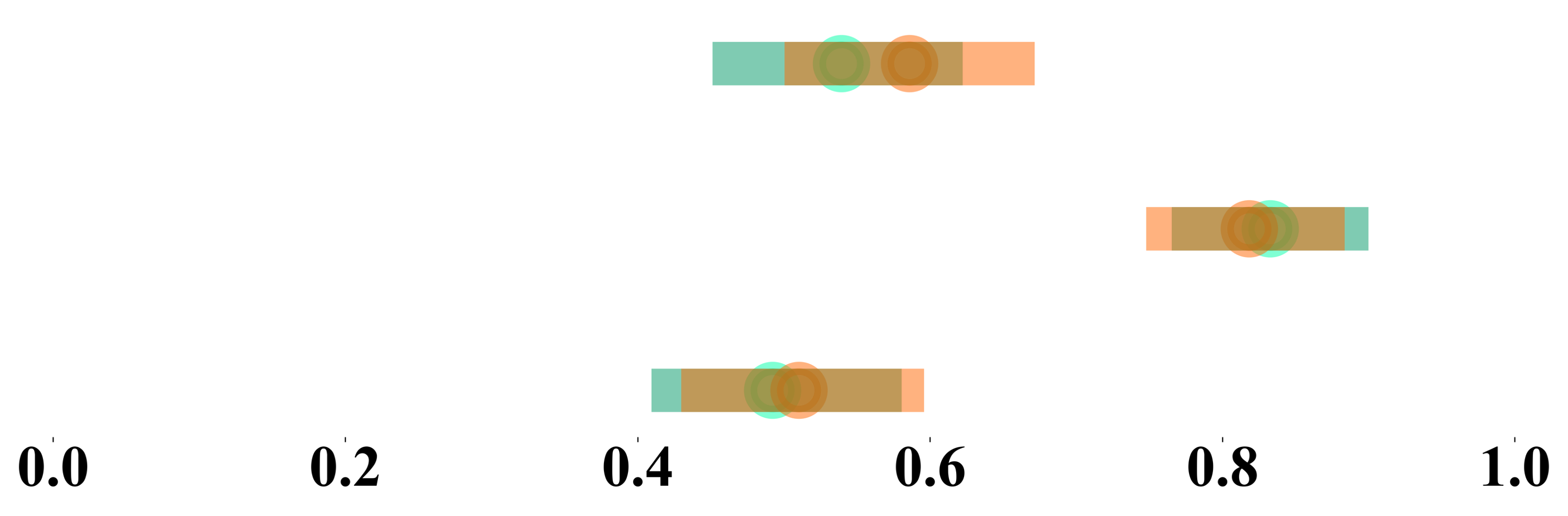}
\hspace*{-3cm}%
}} \\ 
&  &  &  &  &  \\ [-1.5ex]
\hline

%
%
&  &  &  &  &  \\ [-2ex]
18 &  & {\color[HTML]{F19976} About what is the global smartphone market share of Samsung?} & {\color[HTML]{3182BD} 0.47} & 1 &  \\
19 &  & In which company is the global smartphone market share the smallest? & 0.08 & 1 &  \\
20 & \multirow{-3}{*}{Pie Chart} & The global smartphone market share of Apple is larger than that of Huawei. & 0.13 & 0.2 & \multirow{-3}{*}{\raisebox{0pt}[\dimexpr\height-0.4\baselineskip\relax]%
\centering
\hspace*{-.2cm}%
{\includegraphics[width=3.5cm]{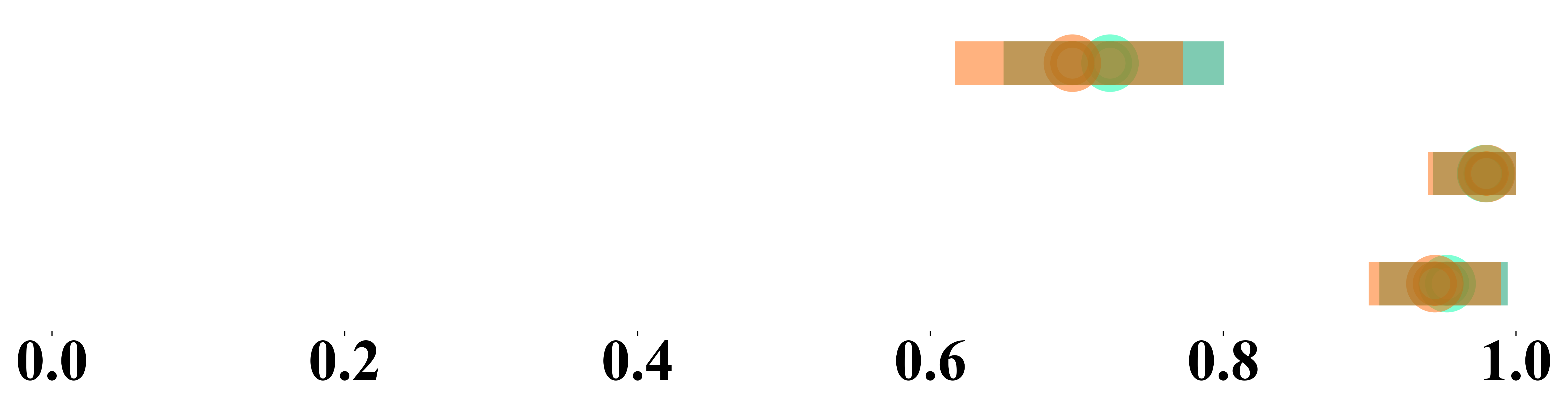}
\hspace*{-3cm}%
}} \\ 
&  &  &  &  &  \\ [-1ex]
\hline

%
%
&  &  &  &  &  \\ [-2ex]
21 &  & How many people have rated the taxi between 4.0 and 4.2? & 0.23 & 0.6 &  \\
22 &  & {\color[HTML]{F19976} What is the rating that the people have rated the taxi the most?} & {\color[HTML]{3182BD} 0.27} & 1 &  \\
23 & \multirow{-3}{*}{Histogram} & More people have rated the taxi between 4.6 and 4.8 than between 4.2 and 4.4. & 0.23 & -0.2 & \multirow{-3}{*}{\raisebox{0pt}[\dimexpr\height-0.4\baselineskip\relax]%
\centering
\hspace*{-.2cm}%
{\includegraphics[width=3.5cm]{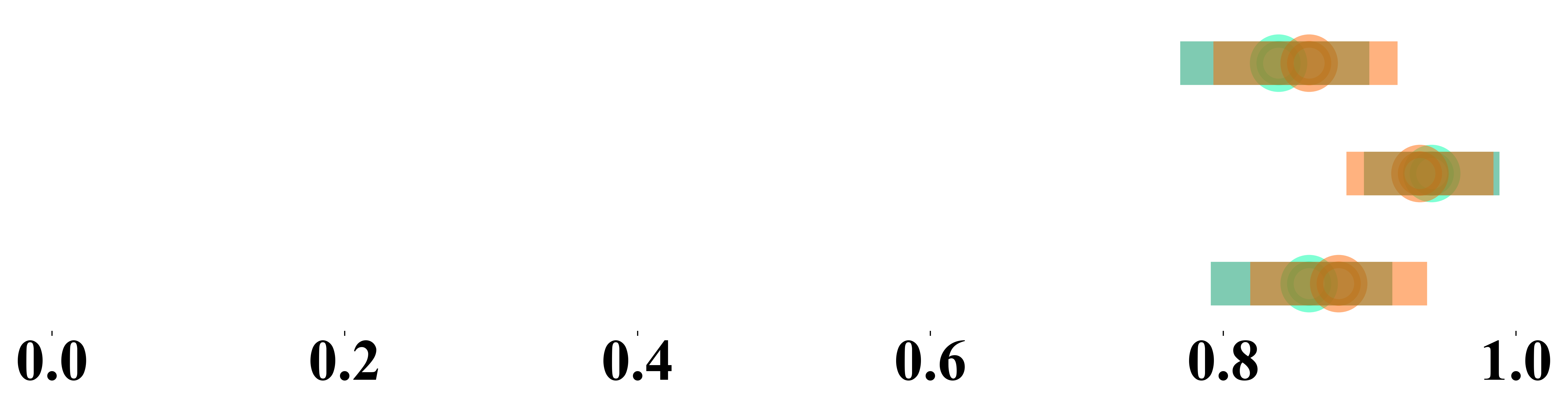}
\hspace*{-3cm}%
}} \\ 
&  &  &  &  &  \\ [-1ex]
\hline

%
%
&  &  &  &  &  \\ [-2ex]
24 &  & What is the weight for the person who is 165.1 cm tall? & 0.41 & 0.2 &  \\
25 &  & What is the height for the tallest person among the 85 males? & {\color[HTML]{3182BD} 0.51} & -0.2 &  \\
26 &  & What is the range in weight for the 85 males? & 0.36 & -0.2 &  \\
27 &  & What is the height for a person who lies outside the others the most? & 0.13 & 0.2 &  \\
28 &  & A group of males are gathered around the height of 176 cm and the weight of 70 kg. & 0.33 & 0.6 &  \\
29 &  & {\color[HTML]{F19976} There is a negative linear relationship between the height and the weight of the 85 males} & 0.43 & 0.6 &  \\
30 & \multirow{-7}{*}{Scatterplot} & The weights for males with the height of 188 cm are all the same. & 0.19 & -0.2 & \multirow{-7}{*}{\raisebox{0pt}[\dimexpr\height-0.3\baselineskip\relax]%
\centering
\hspace*{-.2cm}%
{\includegraphics[width=3.5cm]{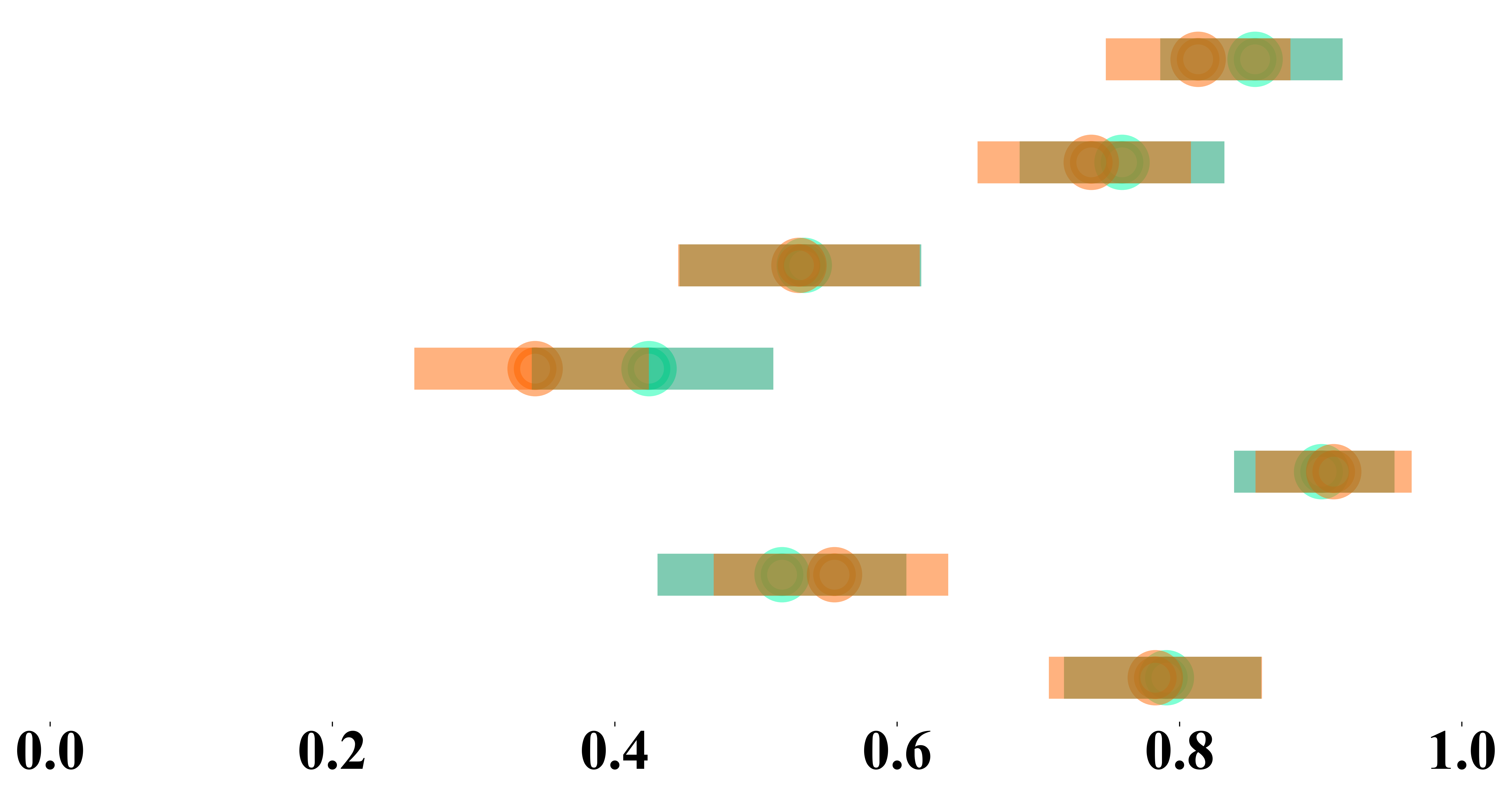}
\hspace*{-3cm}%
}} \\ 
&  &  &  &  &  \\ [-1ex]
\hline

%
%
&  &  &  &  &  \\ [-2ex]
31 &  & {\color[HTML]{F19976} What was the average price of a pound of coffee beans in September 2013?} & {\color[HTML]{3182BD} 0.32} & 0.2 &  \\
32 &  & When was the average price of a pound of coffee beans at minimum? & 0.27 & 0.2 &  \\
33 &  & \begin{tabular}[c]{@{}l@{}}What was the range of the average price of a pound of coffee beans between \\ January 2013 and December 2014?\end{tabular} & 0.17 & -0.6 &  \\
34 & \multirow{-4}{*}{Area Chart} & Over the course of 2013, the average price of a pound of coffee beans was \_\_\_\_\_\_\_\_\_\_\_\_. & 0.27 & -1 & \multirow{-5}{*}{\raisebox{0pt}[\dimexpr\height-0.2\baselineskip\relax]%
\centering
\hspace*{-.2cm}%
{\includegraphics[width=3.5cm]{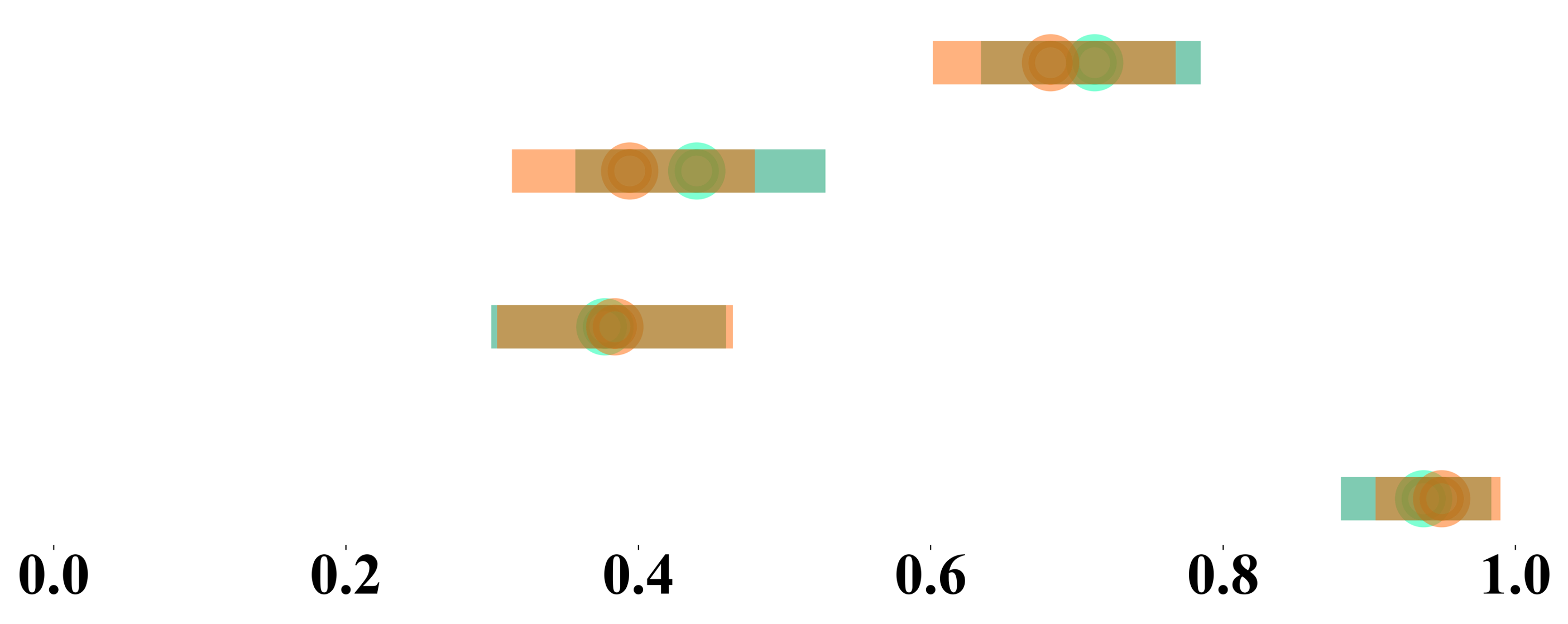}
\hspace*{-3cm}%
}} \\ 
&  &  &  &  &  \\ [-1ex]
\hline

%
%
&  &  &  &  &  \\ [-2ex]
35 &  & What was the number of girls named ‘Amelia’ in 2010 in the UK? & 0.22 & -0.2 &  \\
36 &  & {\color[HTML]{F19976} \begin{tabular}[c]{@{}l@{}}About what was the ratio of the number of girls named ‘Olivia’ to those named \\ ‘Isla’ in 2014 in the UK?\end{tabular}} & {\color[HTML]{3182BD} 0.38} & 0.6 &  \\
37 &  & \begin{tabular}[c]{@{}l@{}}Over the course of years between 2009 and 2014, when was the number of girls \\ named  ‘Amelia’ at the maximum?\end{tabular} & 0.29 & 0.6 &  \\
38 &  & The number of girls named ‘Isla’ was \_\_\_\_\_\_\_\_\_\_ from 2009 to 2012. & 0.18 & -1 &  \\
39 &  & In the UK, the number of girls named ‘Amelia’ in 2014 was more than it was in 2013, & 0.06 & -0.6 &  \\
40 & \multirow{-6}{*}{Stacked Area Chart} & \begin{tabular}[c]{@{}l@{}}Over the course of years between 2009 and 2014, the number of girls named \\ ‘Isla’ was always more than ‘Olivia’.\end{tabular} & 0.23 & 0.2 & \multirow{-8.7}{*}{\raisebox{0pt}[\dimexpr\height-.4\baselineskip\relax]%
\centering
\hspace*{-.2cm}%
{\includegraphics[width=3.5cm]{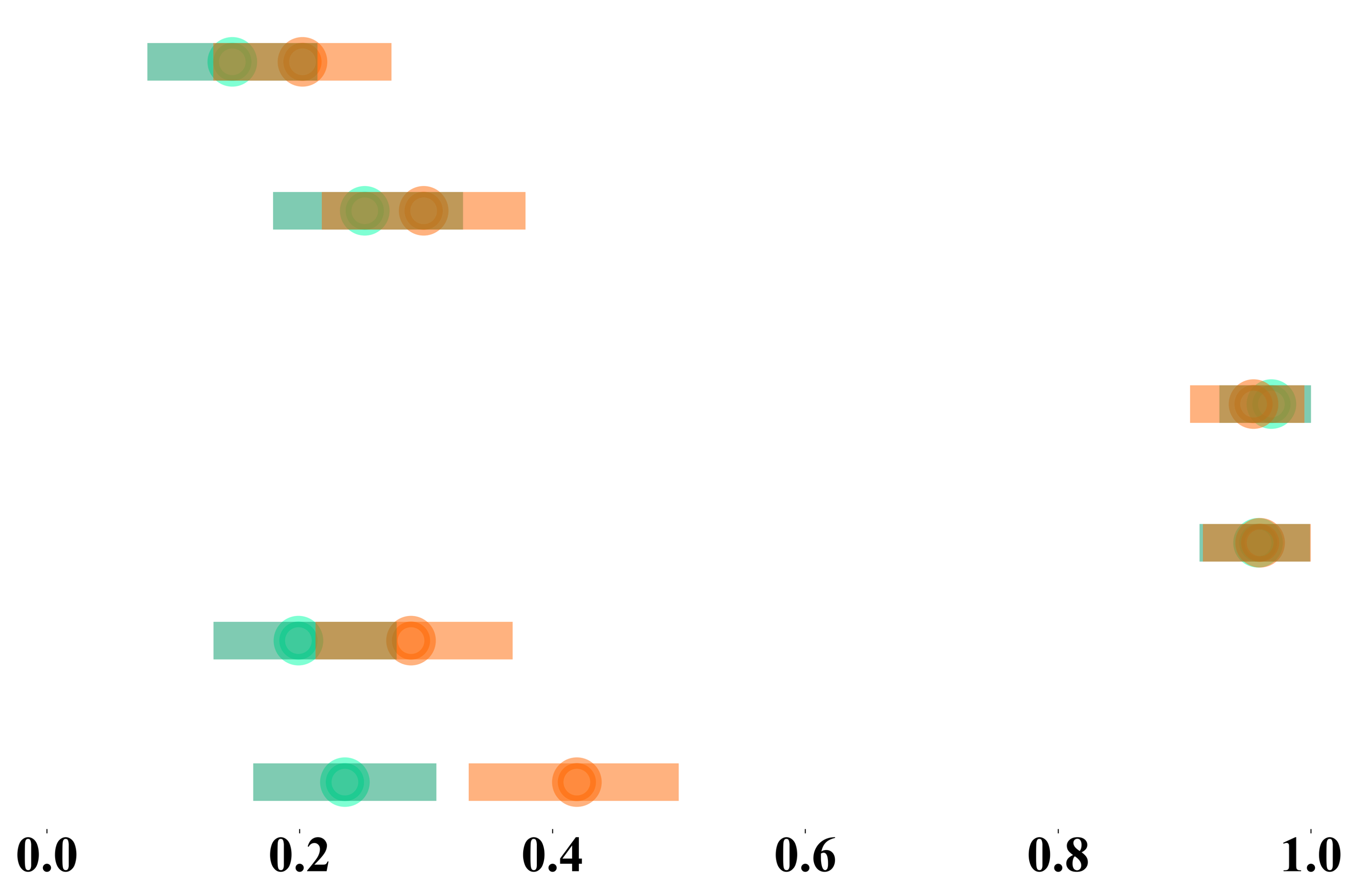}
\hspace*{-3cm}%
}} \\
&  &  &  &  &  \\ [-1ex]
\hline

%
%
&  &  &  &  &  \\ [-2ex]
41 &  & What is the total length of the metro system in Beijing? & 0.15 & -0.2 &  \\
42 &  & {\color[HTML]{F19976} Which city’s metro system has the largest number of stations?} & 0.52 & 0.2 &  \\
43 &  & What is the range of the total length of the metro systems? & {\color[HTML]{3182BD} 0.53} & -0.6 &  \\
44 &  & \begin{tabular}[c]{@{}l@{}}Which city’s metro system does lie outside the relationship between the total system \\ length and the number of stations most?\end{tabular} & 0.38 & 0.2 &  \\
45 &  & \begin{tabular}[c]{@{}l@{}}A group of the metro systems of the world has approximately 300 stations and \\ around a 200 km system length.\end{tabular} & 0.39 & 0.2 &  \\
46 &  & In general, the ridership of the metro system increases as the number of stations increases. & 0.05 & 1 &  \\
47 & \multirow{-7}{*}{Bubble Chart} & The metro system in Shanghai has more ridership than the metro system in Beijing. & 0.48 & 0.2 & \multirow{-9}{*}{\raisebox{0pt}[\dimexpr\height-1.2\baselineskip\relax]%
\centering
\hspace*{-.2cm}%
{\includegraphics[width=3.5cm]{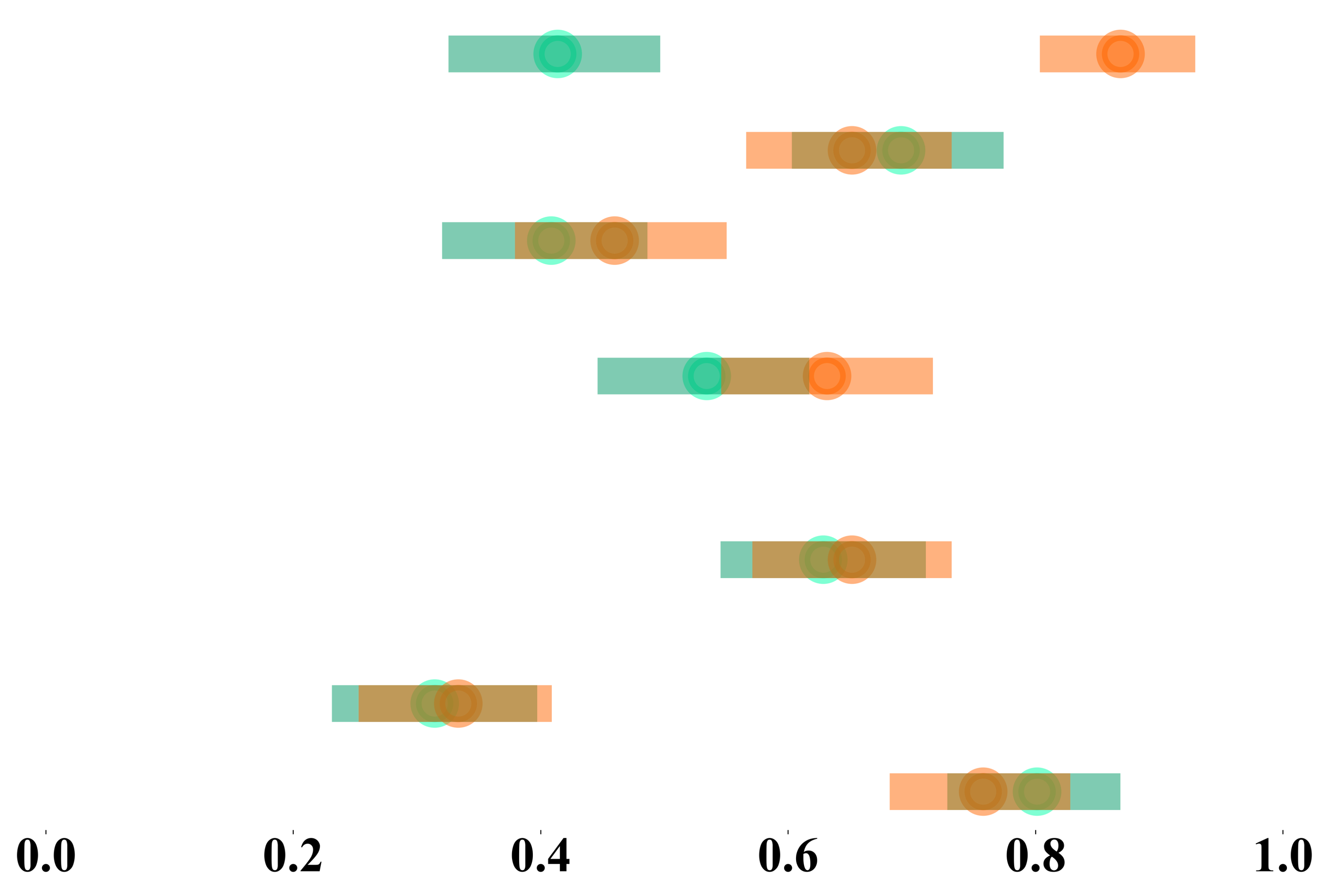}
\hspace*{-3cm}%
}} \\ 
&  &  &  &  &  \\ [-1ex]
\hline

%
%
&  &  &  &  &  \\ [-2ex]
48 &  & What was the unemployment rate for Indiana (IN) in 2015? & 0.24 & 0.2 &  \\
49 &  & In which state was the unemployment rate the highest in 2015? & 0.14 & 0.2 &  \\
50 & \multirow{-3}{*}{Choropleth Map} & {\color[HTML]{F19976} In 2015, the unemployment rate for Washington (WA) was higher than that of Wisconsin (WI).} & {\color[HTML]{3182BD} 0.33} & 0.6 & \multirow{-3}{*}
{\raisebox{0pt}[\dimexpr\height-0.3\baselineskip\relax]%
\centering
\hspace*{-.2cm}%
{\includegraphics[width=3.5cm]{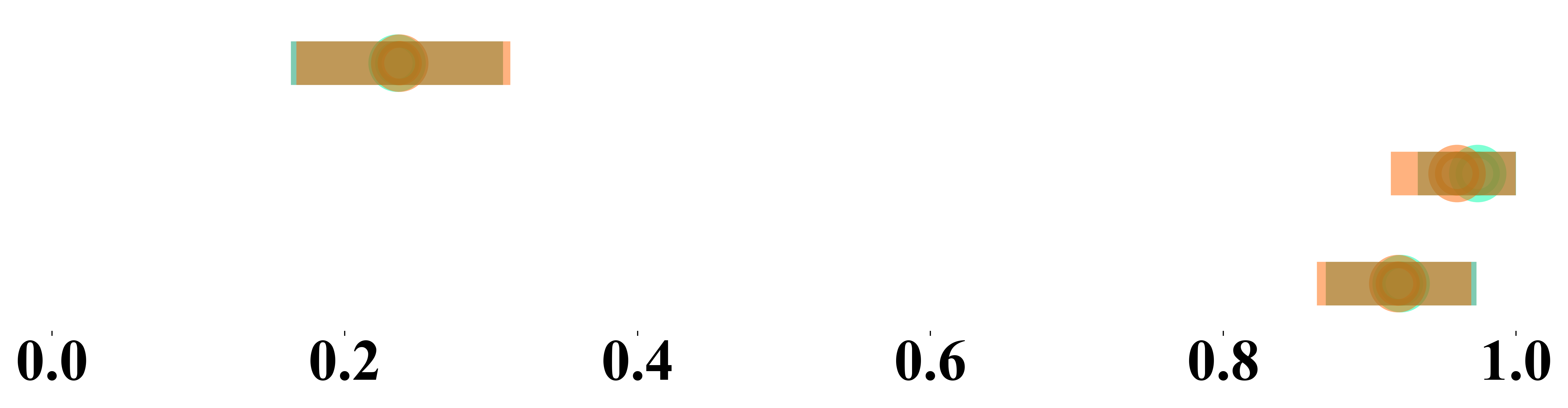}
\hspace*{-3cm}%
}} \\ 
&  &  &  &  &  \\ [-1ex]
\hline

%
%
&  &  &  &  &  \\ [-2ex]
51 &  & For which website was the number of unique visitors the largest in 2010? & 0.21 & 1 &  \\
52 &  & The number of unique visitors for Amazon was more than that of Yahoo in 2010. & 0.29 & 0.6 &  \\
53 & \multirow{-3}{*}{Treemap} & {\color[HTML]{F19976} Samsung is nested in the Financial category.} & {\color[HTML]{3182BD} 0.38} & 1 & \multirow{-3}{*}{\raisebox{0pt}[\dimexpr\height-0\baselineskip\relax]%
\centering
\hspace*{-.2cm}%
{\includegraphics[width=3.5cm]
{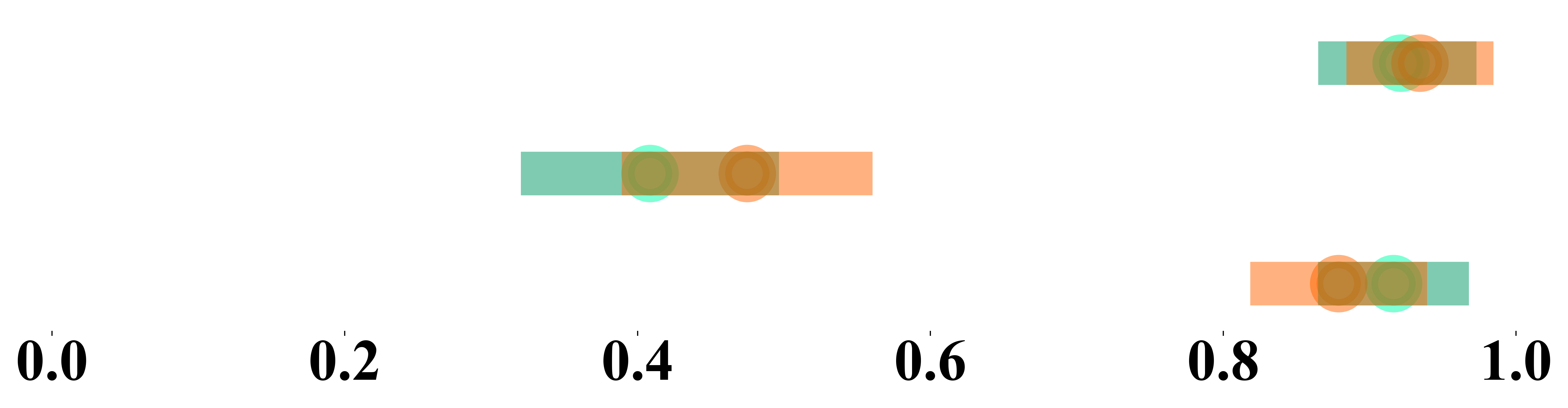}
\hspace*{-3cm}%
}} \\ 
&  &  &  &  &  \\
\hline

\end{tabular*}
\end{table*}

\subsection{Phase I: Test Item Selection}
In this phase, we selected the items for the Mini-VLAT using expert feedback and the item-total correlation. Before choosing the items, we conducted a pilot study to replicate the Mini-VLAT and compute the item-total correlation for each item.

\subsubsection{Replicating VLAT}
\label{replication}
The responses from the original study are not publicly available. As a result, we replicated the VLAT using the images and questions published at \url{https://vlat.herokuapp.com/}. After the replication, we wanted to ensure that the responses obtained in the replicated version significantly overlapped with the original study. Thus, we reached out to the authors of the VLAT, and they shared the data from their pilot study conducted among 191 participants. The horizontal error bars in \autoref{tab:vlat-rep} show that the responses obtained from the {\color[HTML]{FF6600} replication} study for most of the items overlapped with the {\color[HTML]{009966} original} ones. It shows that the questions in the VLAT are strongly reliable, producing stable and consistent results. Therefore, we decided to use the data from the replicated study to select the items for the Mini-VLAT. The pre-registration for this replication study and the data are available at \url{https://osf.io/dy67k/}.

{\large \textbf{Participants}} \hspace{0.3cm}A total of 200 participants were recruited through Prolific for the replication. We applied the same requirements as the participants in VLAT \cite{lee2016vlat}. The participants had a HIT approval rate of 95\% or higher and were limited to the United States. We only accepted participants who were native English speakers. As stated in the original study, we allotted 25 seconds to each question. Out of 200, we only ruled out one participant who was a random clicker. Eventually, a total of 199 participants remained. They comprised 112 females and 85 males with an age range of 19 to 79. Everyone had a high school diploma or higher, 40\% had a bachelor's degree, and 19\% had a master's or a doctoral degree.

{\large \textbf{Scoring}} \hspace{0.3cm} We observed both raw and corrected scores. To calculate the corrected score, we used the correction-for-guessing formula \cite{diamond1973correction, frary1988formula}. Similarly to the original study, the participants were instructed to select the "Skip" option instead of guessing. This setting would influence participants' test-taking strategies and reduce test error caused by guessing, which was a weakness of multiple-choice items \cite{diamond1973correction, frary1988formula}. The correction-for-guessing formula is defined as follows: 
\begin{equation*}
\label{correction-guess}
    CS = R - \frac{W}{C-1}
\end{equation*}
where CS was the corrected score, \textit{R} was the number of items answered correctly (i.e. raw score), \textit{W} was the number of items answered incorrectly, and \textit{C} was the number of choices for an item \cite{lee2016vlat, thorndike2013measurement}

{\large \textbf{Results}} \hspace{0.3cm} The raw scores of the test takers ranged from 18 to 50 (\textit{M} = 35, \textit{SD} = 7.46). The corrected scores ranged from 1 to 48.33 (\textit{M} = 28.13, \textit{SD} = 10.05). In addition, we observed the test completion time. The average test completion time was 11 minutes 33 seconds (\textit{SD} = 2 minutes 17 seconds). The test completion ranged from 5 minutes 45 seconds to 18 minutes 9 seconds. 
After analyzing the responses from the replication study, we computed the item-total correlation for each item, as shown in \autoref{tab:vlat-rep}. In the following sections, we validated the items for the Mini-VLAT with the domain experts using the content validity ratio and computed the item-total correlation for each item in the VLAT.


\subsubsection{Expert Feedback}

In this section, we conducted a content validity evaluation with five domain experts in the visualization community. As the number of experts required for reliable results can be subjective, involving at least five individuals is generally recommended to maintain control over the chance agreement and increase the overall validity of the findings \cite{zamanzadeh2015design, polit2006content}. By having five experts evaluate the relevance of each item to its respective visualization type using a 3-Likert scale, we aimed to ensure that the selected items were appropriate for the Mini-VLAT.

To evaluate content validation, we computed the content validity ratio of each item in the VLAT. The content validity ratio (\textit{CVR}), developed by Lawshe, is a quantitative approach frequently used to assess content validity \cite{lawshe1975quantitative, ayre2014critical}. \textit{CVR} ranges from -1.0 to 1.0 and indicates expert agreement on how a specific item is required to measure the trait or skill. In our evaluation, we are looking for values above zero, indicating that more than half of the panel members agreed upon an item's relevance to the respective visualization. It helps us in item selection by selecting the items with positive \textit{CVR} and high item-total correlation for each visualization type. The outcome of this evaluation would be the first evidence for validity that test items have the meaning intended when the test was developed \cite{mcdonald2013test, mislevy2006implications}.

{\large \textbf{Participants}} \hspace{0.3cm} We invited five domain experts in Information Visualization from various locations. They had 5 to 18 years of experience in visualization research and development. All of the experts were from academia.

{\large \textbf{Procedure}} \hspace{0.3cm} We used the following procedure to evaluate with the experts. First, we informed them about the purpose of the study. Then we asked them to rate each item in the VLAT on a 3-Likert scale, \textit{not necessary, useful but not necessary, and essential}, based on their importance to the respective visualization. We calculated CVR for each item using Lawshe's CVR formula based on the number of experts who indicated "essential" in their responses \cite{ayre2014critical}. 
\begin{equation*}
    CVR = \frac{n_{e} - (N/2)}{N/2}
\end{equation*}
$n_e$ is the number of experts indicating an item as "essential," and $N$ is the total number of experts. 
\autoref{tab:vlat-rep} shows the CVR for each item.

\subsubsection{Item Selection}

Multiple methods exist for selecting short-form items \cite{widaman2011creating}. For example, Lee et al. \cite{lee2016vlat} suggested creating an effective test by selecting highly discriminating items. This approach considers the \textit{discriminating index} of each item, which is defined as the measure of an item's ability to discriminate between those who scored high on the total test and those who scored low on the total test\cite{haladyna2002review}. An alternative approach uses the \textit{item-total correlation}, defined as the correlation between the scores for a given question to the total scores. 

Although there is no consensus for selecting the final items for a test \cite{thorndike2013measurement}, for Mini-VLAT, we chose items with a high item-total correlation. The item-total correlation is more reliable than the items' discriminating index \cite{burton2001item}. It makes maximum use of the available information. In contrast, the discriminating index does not use the data for the middle-scoring group of examinees and only incomplete use of the information for the upper and lower groups \cite{burton2001item}.

After computing the item-total correlation and content validity ratio for each item in the VLAT, we discovered that some of the items (\textit{Item 3}, \textit{Item 8}, \textit{Item 12}, \textit{Item 25}, and \textit{43}) with the highest item-total correlation in their respective visualization type did not have a positive content validity ratio, \textit{CVR}. As a result, we decided to select items with the positive \textit{CVR} and yet high item-total correlation to maintain the validity of the Mini-VLAT, as shown in \autoref{tab:vlat-rep}.

\begin{table*}[ht]
\footnotesize
\centering
\caption{The test items in the Mini-VLAT. The materials used in generating the visualizations (\autoref{fig:mini-vlat-fig}) are available at \url{https://osf.io/46rt8/}.} 
\label{tab:mini-vlat}
\begin{tabular}{|l|l|l|}
\hline
\multicolumn{1}{|c|}{\textbf{Item ID}} & \multicolumn{1}{c|}{\textbf{Visualization Type}} & \multicolumn{1}{c|}{\textbf{Question}} \\ \hline
1 & Line Chart & What was the price of a barrel of oil in February 2020? \\ \hline
2 & Bar Chart & What is the average internet speed in Japan? \\ \hline
3 & Stacked Bar Chart & What is the cost of peanuts in Seoul? \\ \hline
4 & 100\% Stacked Bar Chart & Which country has the lowest proportion of Gold medals? \\ \hline
5 & Pie Chart & What is the approximate global smartphone market share of Samsung? \\ \hline
6 & Histogram & What distance have customers traveled in the taxi the most? \\ \hline
7 & Scatterplot & \begin{tabular}[c]{@{}l@{}}There is a negative linear relationship between the height\\ and the weight of the 85 males.\end{tabular} \\ \hline
8 & Area Chart & What was the average price of pount of coffee beans in October 2019? \\ \hline
9 & Stacked Area Chart & \begin{tabular}[c]{@{}l@{}}What was the ratio of girls named "Isla" to girls\\ named "Amelia" in 2012 in the UK?\end{tabular} \\ \hline
10 & Bubble Chart & Which city’s metro system has the largest number of stations? \\ \hline
11 & Choropleth Map & \begin{tabular}[c]{@{}l@{}}In 2020, the unemployment rate for Washington (WA)\\ was higher than that of Wisconsin (WI).\end{tabular} \\ \hline
12 & Treemap & eBay is nested in the Software category. \\ \hline
\end{tabular}
\end{table*}
\begin{figure*}[h!]
    \centering
    \includegraphics[width=0.9\textwidth]{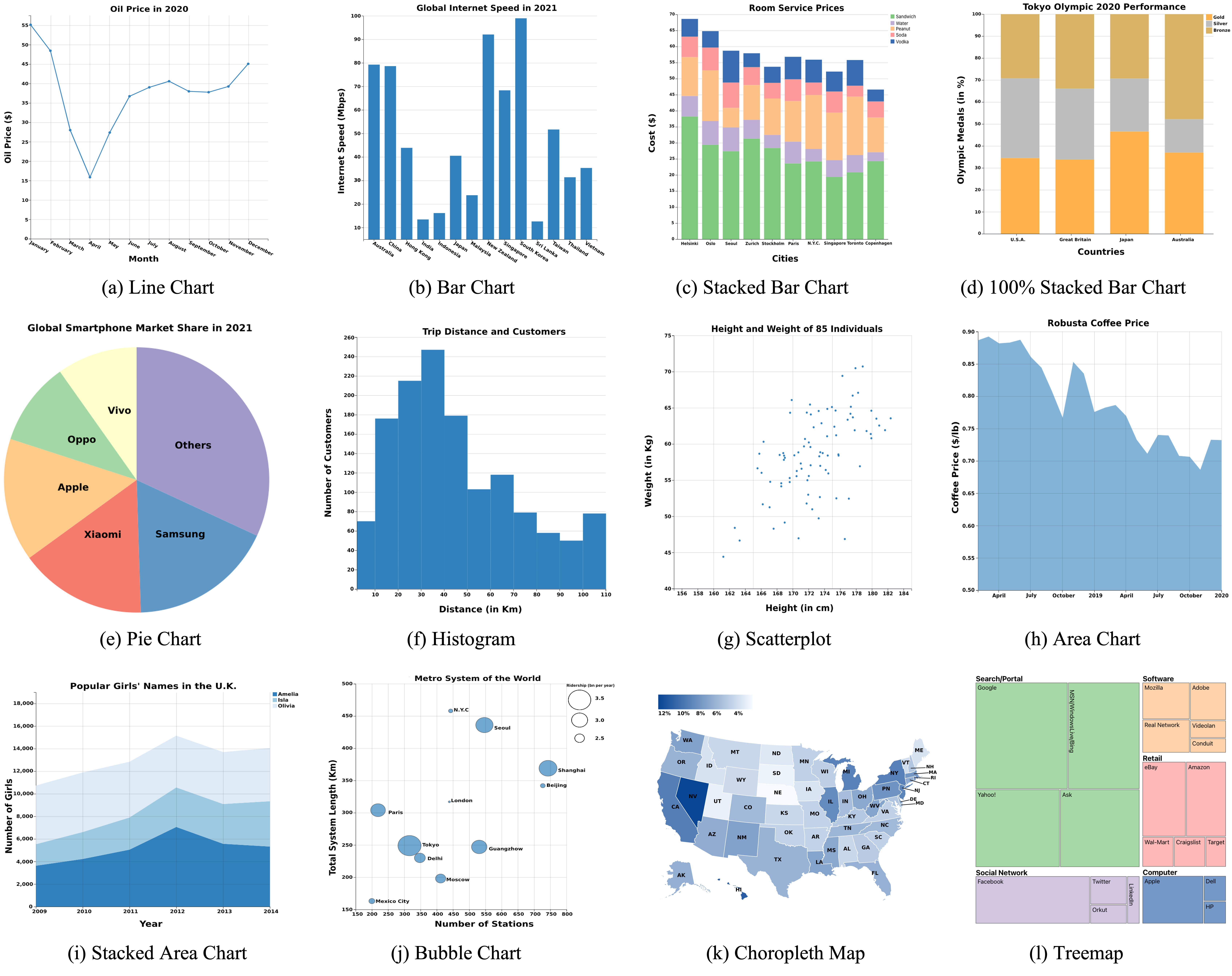}
    \caption{The 12 data visualizations used in the Mini-VLAT.}
    \label{fig:mini-vlat-fig}
    \vspace{1em}
\end{figure*}

\subsection{Phase II: Piloting and Item Refinement}

Once the items were selected, we conducted a pilot study to ensure the test takers easily understood the phrases and vocabulary in the questions. Understanding and assuring the lexical level of potential test takers is critical since the vocabularies and words that make up items might impact test takers' performance \cite{american1999national}. During the pilot study, participants reported difficulty answering questions due to the wording and the unfamiliarity of the topic used in the 100\% Stacked Bar Chart. We then revised the items based on the feedback received in this study. This ensures that the questions in the final Mini-VLAT are not hard to interpret and not inducing any political bias while answering them. 

{\large \textbf{Pilot Study}} \hspace{0.3cm} After selecting the items for the Mini-VLAT, we conducted a pilot study among 15 volunteers via convenience sampling. All of the participants had bachelor's degrees, with 6 of them having completed master's degrees. They consisted of 7 females and 8 males with a self-reported age range of 22 to 59. During the pilot study, 8 participants reported having trouble interpreting some of the questions, and 5 were unfamiliar with the term \textit{exit polls} used in the 100\% Stacked Bar Chart question comparing approval ratings for the Democrat and Republican political parties in the United States. Some reported that questions were too long or required test takers to read at least twice before answering them. 

\noindent
{\large \textbf{Revising Questions}} \hspace{0.3cm} To address the feedback received in the Pilot study, we decided to make several changes to the questions in the Mini-VLAT. We updated the dataset used in the VLAT and created the visualizations using D3.js, \autoref{fig:mini-vlat-fig}. We also used the colorblind safe colors from \textit{ColorBrewer} in creating the visualizations, making it suitable for color-blind users \cite{colorbrew}. For the question regarding 100\% Stacked Bar Chart,  we decided to use the Tokyo 2020 Summer Olympics dataset. The visualization used for 100\% Stacked Bar Chart in the VLAT can induce bias and influence users' responses due to their prior political beliefs or attitude towards a political party \cite{kong2018frames}. The updated and finalized version of the questions in the Mini-VLAT is given in \autoref{tab:mini-vlat}.

\subsection{Phase III: Reliability Evaluation}
After establishing the validity of the Mini-VLAT, we evaluated the reliability of the Mini-VLAT to measure the internal consistency and stability of the test items. Reliability is the property of observed test scores and the attribute of consistency in a test \cite{angoff1953test}. A test's reliability can be measured in several ways, including test-retest reliability, parallel test form reliability, and Cronbach's coefficient alpha \cite{guttman1945basis, lord1983unbiased, peterson1994meta}. Despite the long-standing and widespread use of reliability measures in test development, there does not appear to be a firm consensus on which measure to use \cite{sitarenios2022short}. McDonald's reliability coefficient omega ($\omega$) is another way to calculate the reliability coefficient. It is commonly used in practical situations and is generally preferred over Cronbach's coefficient alpha \cite{hayes2020use, dunn2014alpha}. 

We used the reliability coefficient omega to measure the reliability of the test items in the Mini-VLAT. The result showed that the Mini-VLAT had an acceptable reliability omega of 0.72. This indicates that scores on the test were adequately consistent and were not unduly influenced by random error \cite{nunnally1967psychometric}.

\subsection{Phase IV: Correlation Between Mini-VLAT \& VLAT}
\label{vlat-minivlat-associa}
An essential component of demonstrating the validity of the short form is showing adequate overlapping variance between a short and full form. The short form is analogous to an alternate form of the full-length measure, and a strong correlation between the two forms is required to support the short form's validity. \cite{smith2000sins}. In this section, we performed a crowdsourced experiment where the participants took both the short form and the full form. The pre-registration for Phase IV and V and the data obtained from the crowdsourced experiments are available at \webLink{https://osf.io/46rt8/}.  

{\large \textbf{Participants}} \hspace{0.3cm} We recruited a total of 30 participants using the same requirements as mentioned in \ref{replication} on Prolific. Since the participants were asked to do the VLAT, we excluded those with color blindness. Out of 30, 17 identified as male, and 13 were female. Their ages ranged from 20 to 65. There were no random clickers in this experiment. Everyone had an education level of high school and above, 40\% of the participant had a bachelor's degree, and 6\% of the participants had a master's or a doctoral degree.

{\large \textbf{Procedure}} \hspace{0.3cm} Participants completed two different versions of the assessment: the short-form and full-length. We did not reveal the answers, and we randomly chose whether they saw the short-form and full-length one first. Experimenting with 65 items in one sitting can cause participant fatigue and tiredness, resulting in unreliable scores. Therefore, we provided an optional break of 2 minutes before proceeding to the next test. 

{\large \textbf{Result}} \hspace{0.3cm} We recorded the corrected scores using the correction-for-guessing formula for both Mini-VLAT and VLAT. The corrected scores for Mini-VLAT ranged from -1.33 to 10.67 (\textit{M} = 5.8, \textit{SD} = 2.9) and -10.67 to 45.67 (\textit{M} = 23.3, \textit{SD} = 13.5) for VLAT. In addition, we recorded the completion time for both forms. The completion time for the short form ranged from 2 minutes 21 seconds to 4 minutes 54 seconds (\textit{M}~=~3 minutes 45 seconds, \textit{SD} = 45 seconds ) and 7 minutes 11 seconds to 17 minutes 23 seconds (\textit{M}~=~12 minutes 2 seconds, \textit{SD} = 2 minutes 48 seconds) for the full form.

We calculated a Pearson's product-moment correlation coefficient between the Mini-VLAT and VLAT scores. We noticed a strong positive correlation between the two scores (\textit{r} = 0.75, \textit{n} = 30, \textit{p} < 0.001). Since the correlation was significant, it suggests that the Mini-VLAT is at least as valid as the VLAT. 

\section{PHASE V: TESTING Mini-VLAT'S PREDICTIVE CAPACITY}

Finally, we validated whether Mini-VLAT sufficiently measures someone's ability to read and understand visualizations. A widely accepted approach is \textit{concurrent validity}, which involves
evaluating the predictive capacity of a test score against an independent criterion or a previously validated measure \cite{granpeesheh2014evidence}. An astute reader may note that this is similar to Phase IV. However, given the overlap in the items between VLAT and mini-VLAT, we opted for the extra step of validating with an independent assessment.

Lee et al. \cite{lee2016vlat} demonstrated a strong positive correlation between visualization literacy and the aptitude for learning an unfamiliar visualization using the VLAT scores and the scores on Parallel Coordinate Plot developed by Kwon and Lee \cite{kwon2016comparative}. We used a similar procedure to test whether there was any potential correlation between the Mini-VLAT scores and the results of the well-known Parallel Coordinate Test (P-Lite) to determine whether the meaning of the Mini-VLAT scores could be expanded \cite{firat2022p}.

{\large \textbf{Participants}} \hspace{0.3cm} A total of 30 participants were originally recruited through Prolific. We used similar requirements as mentioned in \ref{replication}. We ruled out 3 random clickers, and none of the participants were aware of Parallel Coordinate Plot in advance. As a result, a total of 27 participants remained. Out of 27, 15 were females, and 12 were males, with a self-reported age range of 22 to 63. Everyone had an education level of high school and above, 41\% of the participant had a bachelor's degree, and 7\% of the participants had a master's or a doctoral degree.

{\large \textbf{Procedure}} \hspace{0.3cm} The experiment was divided into two sections: (1) measuring visualization literacy using the Mini-VLAT and (2) measuring aptitude for learning an unfamiliar visualization. For the first section, we asked participants to answer the 12 items in the Mini-VLAT. After they finished the first section, we introduced them to six static tutorials on Parallel Coordinate Plot before proceeding to the second section \cite{kwon2016comparative}. In the second section, we asked the participants to answer the  14 items obtained from the Parallel Coordinate Plot test, P-Lite \cite{firat2022p}. The questions in both sections were randomized. 

\begin{figure}[h!]
    \centering
    \includegraphics[width=\linewidth]{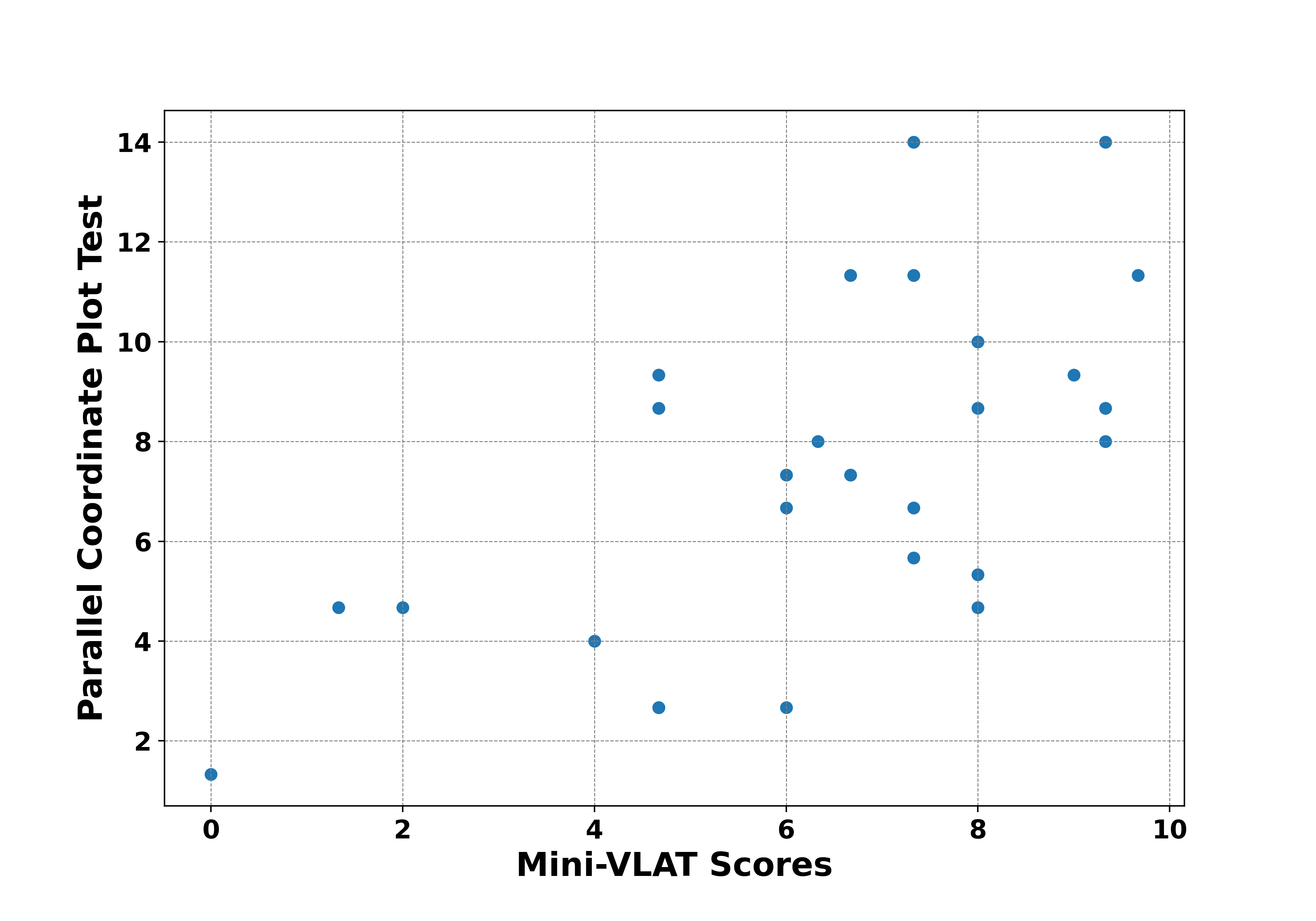}
    \caption{A scatterplot showing the relationship between the Mini-VLAT's scores and the scores on the Parallel Coordinate Test}
    \label{fig:my_label}
\end{figure}

{\large \textbf{Result}} \hspace{0.3cm} 
For the Mini-VLAT, we observed the correct scores ranged from 0 to 9.67 (\textit{M} = 6.36, \textit{S} = 2.49) and 1.33 to 14 (\textit{M} = 7.37, \textit{S} = 3.39) out of 14 for the Parallel Coordinate Test.
We calculated Pearson's product-moment correlation coefficient between the Mini-VLAT scores and the post-tutorial scores to evaluate the relationship between visualization literacy and the aptitude for learning. 
\autoref{fig:my_label} shows a strong positive correlation (\textit{r} = 0.62, \textit{n} = 27, \textit{p} <0.001) between the Mini-VLAT scores and the scores on the aptitude test. This Pearson’s correlation coefficient was reasonably high in Psychological Measurement \cite{schober2018correlation}. This outcome demonstrates a strong and favorable correlation between the Mini-VLAT scores and the users' ability to read and understand unfamiliar visualizations. This strong correlation indicates that the Mini-VLAT has utility in measuring someone's ability to read and understand visualizations

\section{DISCUSSION}

With the increasing use of visualizations, understanding and interpreting the data displayed to formulate accurate observations utilizing visual designs is becoming increasingly important \cite{firat2018towards}. The development of tests, like the one presented in this paper, makes it possible to evaluate users with varying degrees of visualization literacy. However, creating tools/tests to assess visualization literacy is still at the embryonic stage \cite{joshivisualization}.

\subsection{Summary of Results}
This study presents Mini-VLAT, a 12-item short form of the Visualization Literacy Assessment Test (VLAT), a reliable and valid instrument for measuring visualization literacy. The Mini-VLAT was found to have good internal consistency reliability (coefficient omega = 0.72) and content validity (average content validity ratio of 0.6), and it strongly correlates with the VLAT. Additionally, the Mini-VLAT was shown to have a strong positive association with users’ ability to read and understand unfamiliar visualizations.

The results suggest that Mini-VLAT is a practical and time-effective tool for assessing visualization literacy and can be a suitable substitute for the VLAT. The study also highlights the importance of measuring visualization literacy, as understanding and interpreting visualizations are becoming increasingly important in various fields. While there is no agreement on the definition of visualization literacy, this study focused on the ability to read and interpret visualizations, a key aspect of visualization literacy.

Overall, the study provides a foundation for further research on visualization literacy and encourages the development of multiple instruments or exams to assess different aspects of visualization literacy.


\subsection{Toward Measuring Other Dimensions of Visualization Literacy }

Several research studies have attempted to define visualization literacy, but there has been no agreement so far \cite{boy2014principled, borner2016investigating, borner2019data, lee2016vlat, firat2022interactive}. The ability to interpret is included in every visualization literacy definition and accounts for the majority of the literature \cite{solen2022scoping}. However, some previous works have argued that visualization literacy should be defined more broadly as the ability to reason with graphics: knowing when and how to create a visual representation of data to facilitate information extraction and then knowing how to interpret visual representations to read directly from the data \cite{chevalier2018observations, alper2017visualization}. 

It seems unlikely that the research community in the future will agree upon a single definition. As mentioned before, we believe that visualization literacy is a multidimensional construct. This work focused on the user’s \textit{ability to read and interpret data visualizations.} We hope that this work will encourage researchers to investigate the topic of visualization literacy more. In the future, we anticipate that multiple instruments or exams can be developed to assess various aspects of visualization literacy. 

In contrast to previous works primarily focused on the ability to read and interpret well-formed visualizations, a recent study by Lily et al. \cite{lily2023calvi} aimed to measure people's ability to reason about erroneous or potentially misleading visualizations. The researchers developed a precise definition of misleaders and constructed initial test items using a design space of misleaders and chart types. They then tested the provisional assessment on 497 participants and refined the items using various analysis techniques, including Item Response Theory and qualitative analysis. The final bank of 45 items showed high reliability and provided recommendations for future tests and use cases. This study highlights the importance of measuring visualization literacy from a multidimensional perspective and encourages further research.


\section{FUTURE WORK}

Data visualizations have been key in disseminating information to the general public. However, there is still much to do in investigating the visual literacy of the citizens to use the visualizations effectively~\cite{lengler2006identifying, boy2014principled, lee2021viral}. For instance, we need longitudinal studies to understand the ability to comprehend and interpret the visualizations of the general public over a period of time. The Mini-VLAT could assist researchers interested in investigating visualization literacy on a broader scale.

\subsection{Visualization Literacy and Inclusivity}
During our analysis of previous works on accessing and measuring visualization literacy, we noticed a gap in the demographic information of the participant pools during the user studies. Most papers involved participants living in the United States or Europe \cite{lee2016vlat, peppler2021cultivating, alper2017visualization, borner2016investigating, firat2020treemap}.  

However, there have been some works involving those who are largely underrepresented in the data visualization literature. Peck et al. \cite{peck2019data} investigated the 42 community members in rural Pennsylvania about their perceptions of data visualization. Sultana et al. \cite{sultana2021seeing} explored how the visual communication practices in rural tradition differ from the modern communication practices in rural Bangladesh.
The Mini-VLAT can serve as a tool to capture the
visualization literacy of those who are notably neglected in visualization-related studies in a shorter duration of time. 

In addition, previous work has modified the VLAT to investigate a non-English speaking population \cite{kim2021development}. Likewise, a potential future work can be to adopt the Mini-VLAT to different languages, allowing the community to uncover possible cultural differences in how people read and interpret visual data. One way to modify Mini-VLAT to different languages is using machine translation services such as \textit{Google Translate}. This can be a quick and cost-effective method to translate the questionnaire into various languages. However, machine translation services may not always provide accurate translations, especially for idiomatic expressions or technical terms. Therefore, involving native speakers of the target language to validate the translation would be crucial to ensure the accuracy and comprehensibility of the Mini-VLAT in different languages. Further Modifying Mini-VLAT to different languages can provide valuable insights into the visualization literacy and interpretation of visual representations among diverse populations.

\section{LIMITATION}
Although we strictly followed the best practices for creating a short form, several noteworthy limitations exist. For example, the Mini-VLAT is limited to 12 different visualization types. We adopted the visualizations from the VLAT, which surveyed the most frequently occurring visualization types in news outlets and the K-12 curriculum \cite{lee2016vlat}. This assessment does not represent many other common data representations. For example, it would be beneficial to include icon arrays and other statistical charts, often used for medical decision-making~\cite{micallef2012assessing,ottley2012visually,ottley2019curious,ottley2015improving,mosca2021does}. 
Other charts, such as radar charts and heat maps, are notably excluded. Thus, future work is necessary to expand the range of visualization types included in the Mini-VLAT, ensuring that the assessments cover a more comprehensive range of visualizations.

As mentioned by Lee et al.~\cite{lee2016vlat}, it could be argued that visualization literacy is too complex to assign a number. Laura M. Ahearn \cite{ahearn2004literacy} described literacy as multiple rather than unidirectional, focusing on the effect of local conditions on how individuals of the community practice and perceive the significance of literacy. Arneson et al. \cite{arneson2018visual} proposed the Visualization Blooming Tool (VBT), an adaption of Bloom's taxonomy, to evaluate scientific visual literacy in undergraduate instruction. Evaluating visual literacy based on this taxonomy might require expert intervention when accessing visual literacy and might not be suitable for large-scale experiments. One of the solutions could be to develop tools or tests that cover the individual stage of the cognitive taxonomy mentioned in the VBT.

One might also argue that the length of the Mini-VLAT is too short, and the exact number of items was discretionary. Nonetheless, the reliability and validity of the shorter 12-item Mini-VLAT are almost as strong as those of the full 53-item VLAT.
Since creating tools/tests to access visual literacy is still in the early stage, we believe that having an efficient and validated test is an important step toward investigating visualization literacy \cite{lee2016vlat}. Our findings suggest that the Mini-VLAT is a viable tool for researchers and scholars interested in effectively evaluating visualization literacy. Since this is based on the empirical results presented here, we want to emphasize that researchers should consider the advantages and disadvantages of adopting abbreviated forms \cite{donnellan2006mini}. We believe it is important to use a practical perspective while making this decision, namely that of test participants who are frequently faced with the difficulty of completing long surveys.

\section{CONCLUSION}
In this paper, we systematically developed an efficient and validated tool, Mini-VLAT, to measure visualization literacy. We also demonstrated the predictive capacity of the Mini-VLAT towards the aptitude for learning on unfamiliar visualizations. 
We believe that the Mini-VLAT will help the visualization community evaluate visualization literacy on a broader scale and encourage researchers to understand users’ visual literacy better and more efficiently. In addition, we have demonstrated how to redesign or shorten a test while maintaining acceptable internal consistency and validity to measure visualization literacy. We intend to build on this work to access the visualization literacy of diverse populations. 

\section{ACKNOWLEDGMENTS}
We acknowledge Melanie Bancilhon for her constructive advice when selecting and revising items for the Mini-VLAT. We thank the participants in the pilot study for helping us refine the items in the Mini-VLAT and the experts participating in the content validity evaluation. Lastly, we also thank Bum Chul Kwon and Sung-Hee Kim for sharing the data from the original pilot study and discussing the insights from their prior work. This material is based upon work supported by the National Science Foundation under grant number IIS-2142977. 

\bibliographystyle{eg-alpha-doi}
\bibliography{egbibsample}



\end{document}